# Droplet impact on a superhydrophobic surface under shear airflow: Lattice Boltzmann simulations and scaling analyses

Yang Liu[1], Xuan Zhang[2], Yiqing Guo[1], Xiaomin Wu[1, *], Jingchun Min[3, *]

[1] Key Laboratory for Thermal Science and Power Engineering of Ministry of Education,

Department of Energy and Power Engineering, Tsinghua University, Beijing 100084, China

[2] Department of Energy and Power Engineering, School of Mechanical Engineering,

Beijing Institute of Technology, Beijing 100081, China

[3] Key Laboratory for Thermal Science and Power Engineering of Ministry of Education,

Department of Engineering Mechanics, Tsinghua University, Beijing 100084, China

*Corresponding author(s): wuxiaomin@mail.tsinghua.edu.cn (X.M. Wu), minjc@mail.tsinghua.edu.cn (J.C. Min)

**Abstract:** Droplet impact in airflow environments is ubiquitous in nature and industry, making the understanding of this multiphase behavior crucial for technologies such as anti-icing and spray cooling. In this study, the dynamics of droplet impact on a superhydrophobic surface under shear airflow are numerically investigated using the pseudopotential multiphase lattice Boltzmann method. This three-dimensional model employs a non-orthogonal multiple-relaxation-time scheme to enhance numerical stability and a contact angle hysteresis window to effectively capture dynamic wetting. Specifically, the kinetic energy supplied by the airflow enhances streamwise spreading and significantly expands the final contact footprint due to continuous horizontal sliding. To describe the nonlinear dependence of these contact-line characteristics on the impact Weber number ($We$) and the airflow Reynolds number ($Re$), a set of composite scaling laws is developed based on a modified Weber number ($We^*$) that incorporates the airflow contribution. Moreover, the aerodynamic effect leads to a higher velocity restitution coefficient and a deflected take-off angle. Based on an energy partition analysis at detachment, a refined power law is derived to scale the vertical restitution coefficient under shear airflow, while the streamwise restitution coefficient is formulated via the sliding velocity approximation. Integrating these two directional components enables accurate quantitative predictions of the total restitution coefficient and the take-off angle governed by the interplay of $We$ and $Re$. Overall, this study clarifies the underlying mechanisms of droplet-airflow-surface interactions, providing practical insights for predicting droplet behaviors and guiding surface design under aerodynamic conditions.

## 1. Introduction

Falling droplets and airflow are ubiquitous fluid phenomena in nature, and their interplay is central to numerous engineering applications, such as ship or aircraft transportation, wind power generation, power transmission via cables, and spray cooling [1-3]. In recent years, experimental imaging and numerical simulations have provided significant insights into the mechanism of this gas-liquid interaction behavior, particularly regarding interfacial instabilities like droplet deformation, breakup, and atomization in free-space flow fields [4-9]. However, the droplet impact process remains underexplored in confined scenarios where shear flow, droplets, and solid surfaces coexist. In this context, superhydrophobic surfaces (SHPS) have attracted widespread attention for practical applications like anti-icing and self-cleaning due to their extremely low adhesion and excellent rebound characteristics [10].

Recent research on the dynamic behavior of droplet impact on SHPS has made substantial progress. Extensive studies have characterized key aspects such as spreading diameter [11, 12], contact time [13, 14], impact force [15, 16], and rebound characteristics [17, 18]. Furthermore, the regulating effects of surface elasticity [19, 20], movement speed [21, 22], and inclination angle [23, 24], and supercooling degree [25, 26] have been systematically examined. These works enrich our knowledge of how inertia, viscosity, and surface tension govern the spreading and retraction processes on SHPS, thereby facilitating the development of droplet manipulation techniques and high-performance surface or coating fabrication strategies [27]. Nevertheless, most of these experimental studies were conducted in quiescent laboratory environments, overlooking the tangential effects of shear airflow. Although droplet impingement on SHPS typically lasts only milliseconds, the surface's low contact angle hysteresis and high interfacial mobility render the asymmetric influence of airflow non-negligible. Consequently, the investigation of impact dynamics in the context of shear airflow is imperative for comprehending droplet-surface interactions within realistic environments and optimising surface design to enhance functional performance.

As for the droplet dynamics behavior on surfaces under shear flow, most studies to date have primarily focused on forced phenomena such as droplet morphology, sliding or rolling, and detachment state. Ding et al. [28] employed a dissipative interface method to simulate the shear flow influence on the contact-line motion of a sessile droplet. Their results identified multiple regimes of droplet motion under the control of capillary number and gas-liquid viscosity ratio, containing steady sliding, partial pinching-off, and complete detachment. Ruyer-Quil et al. [29], using a shallow-water equation model incorporating capillary effects and vorticity tensors, numerically demonstrated that contact angle hysteresis not only determines whether a droplet slides in a teardrop-like or bullet-shaped form, but also significantly reduces its sliding speed and delays coalescence. Wang et al. [30] experimentally characterized various wetting morphologies of sessile droplets under airflow using high-speed imaging, and systematically quantified how airflow velocity, droplet size, and surface wettability affect the dynamic contact angle, slide displacement and speed. Except for airflow and waterdrop, Shang et al. [31] focused on shear solute flows containing surfactants and investigated how concentration gradients influence droplet motion and detachment behavior induced by Marangoni stress, offering valuable insight into surfactant-mediated droplet manipulation. Also, Chahine et al. [32] reported a unique caterpillar-like sliding mode for glycerol droplets, characterized by a contraction-extension motion, which ultimately leads to droplet fragmentation once a critical stretch is reached, followed by unstable coalescence. Collectively, the existing studies elucidate the nonlinear dynamics of sessile droplets subjected to shear flow, contributing to a better understanding of this commonly encountered multiphase system.

Interestingly, the interplay between airflow, droplets, and surfaces can also produce a special flow phenomenon, namely the aerodynamic Leidenfrost effect [33, 34]. On this basis, Yu et al. [35] proposed a novel strategy for directional droplet rebound without requiring surface wettability modification, which forms an airflow-induced gas film upon the surface to repel away the impinging droplet by aerodynamic lift. However, to the best of our knowledge, research on the coexistence of shear airflow, impinging droplet, and SHPS remain scarce, with limited mechanistic analysis and quantitative characterization for their coupling behaviors [36].

To this end, a three-dimensional lattice Boltzmann (LB) method is employed to numerically investigate the droplet impact dynamics on SHPS under shear airflow. The remainder of this paper is organized as follows. First, the multiphase numerical framework, physical model, and comprehensive experimental validations under both quiescent and airflow conditions are introduced. Next, the coupled effects of airflow velocity and impact inertia on droplet morphology and contact-line behavior are systematically explored. Based on a modified Weber number incorporating the kinetic energy contribution of the airflow, a set of theoretical scaling laws is developed to quantitatively predict the spreading factor, slide displacement, and the final contact footprint. Finally, the energy conversion at detachment is examined to derive analytical models for the velocity restitution coefficients and the take-off angle, thereby providing a comprehensive kinematic description of the droplet rebound behavior.

## 2. Methodology and validation

### 2.1. Pseudopotential multiphase lattice Boltzmann model

As an emerging mesoscale technology, the LB method features several prominent advantages in flow simulation, including the ability to naturally capture topological changes at the liquid-gas interface without explicit tracking, ease of modeling multiphase systems, and high efficiency in parallel computation [37-40]. In this work, the pseudopotential LB model is employed as the multiphase solver, with the non-orthogonal multiple-relaxation-time (NMRT) collision operator adopted to enhance numerical stability in simulating two-phase systems with high density ratios [41, 42]. The single-field LB evolution equation governing the liquid-gas flow reads [43, 44]

$$f_i(\mathbf{x}+\mathbf{e}_i\Delta t, t+\Delta t) = f_i(\mathbf{x},t) - \mathbf{M}^{-1}\mathbf{\Gamma}(\mathbf{m}_i - \mathbf{m}_i^{eq}) + \mathbf{M}^{-1}(\mathbf{I} - \mathbf{\Gamma}/2)\mathbf{S} \quad (1)$$

where $f_i$ represents the density distribution function in the $i$-th discrete velocity at position $\mathbf{x}$ and time $t$, whose moment-space form is represented as $\mathbf{m}=\mathbf{Mf}$. The superscript $eq$ denotes the equilibrium distribution, $\Delta t$ is the time step, $\mathbf{I}$ is the identity matrix, and $\mathbf{S}$ denotes the discrete forcing term in moment space. For the D3Q19 lattice stencil, the discrete velocity vectors $\mathbf{e}_i$ and the non-orthogonal transformation matrix $\mathbf{M}$ are detailed in the *Appendix*. The diagonal relaxation matrix $\mathbf{\Gamma}$ for the NMRT scheme is chosen as [45]

$$\boldsymbol{\Gamma} = \mathrm{diag}\left(1,1,1,1,s_v,s_v,s_v,s_b,s_v,s_v,s_v,s_3,s_3,s_3,s_3,s_3,s_3,s_4,s_4,s_4\right) \tag{2}$$

where $s_v$ and $s_b$ are related to the fluid kinematic viscosity $v=(1/s_v-0.5)c_s^2\Delta t$ and bulk viscosity $v_b=2/3(1/s_b-0.5)c_s^2\Delta t$, therein $c_s=c/\sqrt{3}$ is the lattice sound speed, and the lattice velocity is fixed at $c=\Delta x/\Delta t=1$ in this work. The other relaxation parameters are chosen as $s_b$=0.8, $s_3$=$s_4$=1.2 for simulation stability. The equilibrium moment $\mathbf{m}^{eq}$ is explicitly given by

$$\begin{aligned}\mathbf{m}^{eq}=[&\rho,\rho u_x,\rho u_y,\rho u_z,\rho u_x u_y,\rho u_x u_z,\rho u_y u_z,\rho(1+\mathbf{u}^2),\rho(u_x^2-u_y^2),\rho(u_x^2-u_z^2),\rho c_s^2 u_x,\rho c_s^2 u_x,\\ &\rho c_s^2 u_y,\rho c_s^2 u_z,\rho c_s^2 u_y,\rho c_s^2 u_z,\rho c_s^2(c_s^2+u_x^2+u_y^2),\rho c_s^2(c_s^2+u_x^2+u_z^2),\rho c_s^2(c_s^2+u_y^2+u_z^2)]^{\mathrm{T}}\end{aligned} \tag{3}$$

where $\rho$ and $\mathbf{u}$ represent the fluid density and macroscopic velocity, respectively, calculated as

$$\rho=\sum_i f_i,\ \rho\mathbf{u}=\sum_i f_i\mathbf{e}_i+\frac{\Delta t}{2}\mathbf{F} \tag{4}$$

where $\mathbf{F}=\mathbf{F}_{int}+\rho g$ denotes the total force exerted on fluid, therein $g$ is gravitational acceleration, and $\mathbf{F}_{int}$ represents the two-phase interaction force, given by

$$\mathbf{F}_{int}=-G_{int}\psi(\mathbf{x})\sum_i \omega(|\mathbf{e}_i|^2)\psi(\mathbf{x}+\mathbf{e}_i)\mathbf{e}_i c_s^2 \tag{5}$$

where $\mathbf{G}_{int}$= -1 is a constant for the interaction strength between phases; $\omega(|\mathbf{e}_i|^2)$ are the weight coefficients in the D3Q19 model, with $\omega(0)$=1/3, $\omega(1)$=1/18, $\omega(2)$=1/36. The pseudopotential function $\psi$ is defined as

$$\psi(\mathbf{x})=\sqrt{2(P_{EOS}-\rho c_s^2)/(Gc^2)} \tag{6}$$

where $P_{EOS}$ is the liquid-gas fluid pressure, and the piecewise linear equation of state is incorporated here, responsible for the separation of two phases with high density ratios [42, 43], i.e.,

$$P_{EOS}(\rho)=\begin{cases}\rho\theta_g, \rho\le\rho_1\\ \rho_1\theta_g+(\rho-\rho_1)\theta_m, \rho_1\le\rho\le\rho_2\\ \rho_1\theta_g+(\rho_2-\rho_1)\theta_m+(\rho-\rho_2)\theta_l, \rho>\rho_2\end{cases} \tag{7}$$

in which $\theta_g=c_s^2/2, \theta_l=c_s^2, \theta_m=-c_s^2/40$ are chosen as the equation parameters to reproduce the coexistence of liquid water and gaseous air, and the spinodal points $\rho_1$=0.001325, $\rho_2$=0.9758 are determined based on mechanical stability and thermodynamic equilibrium conditions.

Furthermore, to achieve thermodynamic consistency for the water-air two-phase system, the improved discrete force scheme proposed by Li et al. [46] is adopted, i.e.,

$$\begin{aligned}\mathbf{S}=[&0,F_x,F_y,F_z,F_x u_y+F_y u_x-\frac{Q_{xy}}{3v},F_x u_z+F_z u_x-\frac{Q_{xz}}{3v},F_y u_z+F_z u_y-\frac{Q_{yz}}{3v},\\ &2\mathbf{F}\cdot\mathbf{u}+\frac{4\epsilon|\mathbf{F}_{int}|^2}{3v_b\psi^2}+\frac{8(Q_{xx}+Q_{yy}+Q_{zz})}{45v_b},2(F_x u_x-F_y u_y)-\frac{Q_{xx}-Q_{yy}}{3v},2(F_x u_x-F_z u_z)-\frac{Q_{xx}-Q_{zz}}{3v},\\ &F_x c_s^2,F_x c_s^2,F_y c_s^2,F_z c_s^2,F_y c_s^2,F_z c_s^2,2c_s^2(F_x u_x+F_y u_y),2c_s^2(F_x u_x+F_z u_z),2c_s^2(F_y u_y+F_z u_z)]^{\mathrm{T}}\end{aligned} \tag{8}$$

where the model parameter $\epsilon$ is fixed at 0.1. To ensure the accuracy of multiphase simulation, an additional correction term $Q_{\alpha\beta}$ is introduced, calculated via[47]

$$Q_{\alpha\beta}=\kappa\frac{G_{int}}{2}\psi(\mathbf{x})\sum_i \omega(|\mathbf{e}_i|^2)(\psi(\mathbf{x}+\mathbf{e}_i)-\psi(\mathbf{x}))\mathbf{e}_i\mathbf{e}_i \tag{9}$$

in which $\kappa$ is a constant used to adjust the surface tension between two phases, with its value examined in the *Appendix*.

## 2.2. Computational domain and boundary conditions

The three-dimensional model for droplet impact on SHPS under shear airflow is illustrated in Fig. 1(a), while the quantitative parameters characterizing the droplet behavior are labeled in Fig. 1(b). The computational domain occupies the space of $L \times H \times H$, with periodic boundary conditions along the $y$-direction. Initially, the impact velocity of the droplet with diameter $D_0$ is prescribed as $u_{d,z}$, and its gravity center is positioned in the $xz$-plane ($y=H/2$) at a distance of $D_0$ from both the upstream airflow inlet and the bottom solid wall. During the spreading and retraction stages, the spreading diameters on the contact surface along the streamwise ($x$) and transverse ($y$) directions are $D_x$ and $D_y$, respectively, while the maximum droplet widths are denoted as $W_x$ and $W_y$. The two red symbols marked in Fig. 1(b) indicate the initial impact and detachment points of the droplet, the horizontal distance between them is defined as the slide displacement $L_x$, and the take-off angle at the detachment moment is denoted as $\alpha$.

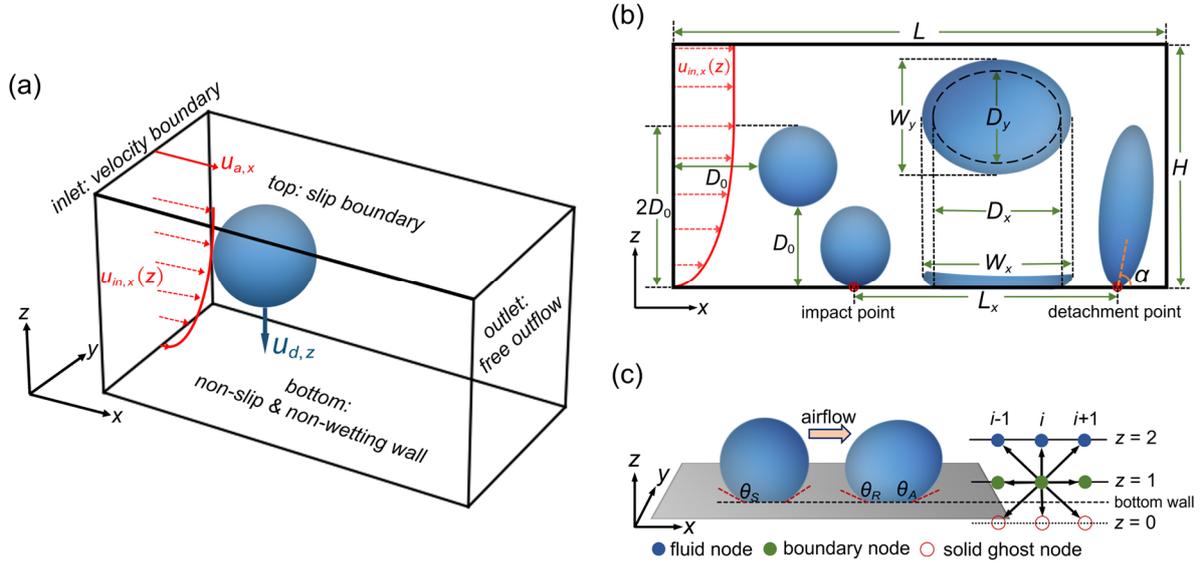

**Fig. 1.** Numerical setup of droplet impact on the SHPS under shear airflow: (a) Schematic diagram of the 3D computational model; (b) Definition of the quantitative parameters during droplet spreading, retraction, sliding, and detachment; (c) Illustration of the implementation of contact angle hysteresis window.

As highlighted by the red line in Fig. 1 (a) and (b), the $+x$ direction velocity $u_{in,x}$ is treated as a piecewise parabolic profile to emulate a shear flow field within a boundary layer [48, 49], i.e.,

$$u_{in,x}(z) = \begin{cases} u_{a,x} \dfrac{z}{4D_0^2}(4D_0 - z), & 0 < z < 2D_0 \\ u_{a,x}, & 2D_0 \leq z < H \end{cases} \quad (10)$$

where $u_{a,x}$ is the mainstream airflow velocity, and the airflow layer ($0<z<2D_0$) is aligned with the droplet apex in height, and the inlet condition is implemented using Zou-He velocity boundary scheme [50, 51], i.e.,

$$f_{-i}(\mathbf{x}, t+\Delta t) = f_i(\mathbf{x}, t) - \frac{\rho}{6}\mathbf{e}_i \cdot \mathbf{u}_{in} - \frac{\rho}{3}\mathbf{t}_i \cdot \mathbf{u}_{in} + \frac{1}{2}\sum_{j=1}^{19} f_j(\mathbf{t}_i \cdot \mathbf{e}_j)(1 - |\mathbf{e}_j \cdot \mathbf{n}_{in}|) \quad (11)$$

where $f_i$ and $f_{-i}$ represent the distribution functions directed outward and inward the inlet ($\mathbf{e}_{-i} = -\mathbf{e}_i$); $\mathbf{n}_{in} = (1,0,0)$ and $\mathbf{t}_i = \mathbf{e}_i - (\mathbf{e}_i \cdot \mathbf{n}_{in})\mathbf{n}_{in}$ are the unit normal and tangential vectors of the inlet, respectively; $\mathbf{u}_{in} = (\mathbf{u}_{in,x}, 0, 0)$ is the inlet velocity vector. At the top boundary, the slip vector $\mathbf{u}_a = (\mathbf{u}_{a,x}, 0, 0)$ is imposed to signify the bulk region, and this Dirichlet condition is realized via the bounce back scheme

$$f_{-i}(\mathbf{x}, t+\Delta t) = f_i(\mathbf{x}, t) - 2\omega_i \rho \, \mathbf{e}_i \cdot \mathbf{u}_a / c_s^2 \tag{12}$$

The domain outlet is arranged as a free outflow boundary, adopting the non-equilibrium extrapolation scheme [52]

$$f_i(\mathbf{x}, t+\Delta t) = f_i(\mathbf{x}^*, t) + \left( f_i^{eq}(\rho, \mathbf{u}^*) - f_i^{eq}(\rho^*, \mathbf{u}^*) \right) \tag{13}$$

where the superscript * denotes the nearest interior lattice node to the outlet.

Moreover, the bottom boundary is modeled as the no-slip SHPS. To minimize spurious currents and precisely control wettability, the geometric wetting boundary scheme is employed. As illustrated in Fig. 1(c), the virtual fluid density at the solid ghost node is determined via the geometric relation of the prescribed contact angle $\theta$ using the following discretized differential calculation [53]

$$\rho_{i,j,0} = \rho_{i,j,2} + \tan\left(\frac{\pi}{2} - \theta\right) \xi$$
$$\xi = \sqrt{(\rho_{i+1,j,1} - \rho_{i-1,j,1})^2 + (\rho_{i,j+1,1} - \rho_{i,j-1,1})^2} \tag{14}$$

where the subscripts $i$, $j$, and $z$ denote the spatial coordinates. Furthermore, to effectively capture the dynamic wetting behavior of the droplet under shear airflow, the contact angle hysteresis model is implemented [54-56]. The hysteresis window is prescribed as $\theta_R \leq \theta \leq \theta_A$, with the advancing angle $\theta_A = 155°$ and the receding angle $\theta_R = 145°$. These values are selected to represent a generic SHPS with the static contact angle $\theta_S = 150°$ [57]. For numerical implementation, the local contact angle $\theta$ at each time step is estimated by inversely solving Eq. (14) from the virtual solid density of the previous time step. If $\theta < \theta_R$, $\theta$ is readjusted as $\theta_R$; if $\theta > \theta_A$, $\theta$ is replaced by $\theta_A$. After executing the above procedures, $\theta$ is then re-substituted into Eq. (14) to update the virtual density for the next time step, thus reproducing the hysteresis of contact-line movement.

*2.3. Model validation*

This subsection validates the multiphase model and boundary conditions introduced in Secs. 2.1 and 2.2. The densities and kinematic viscosities of the liquid-gas fluids are set as $\rho_l = 1000$ kg/m$^3$, $\rho_g = 1.0$ kg/m$^3$, $\upsilon_l = 1.0 \times 10^{-6}$ m$^2$/s, and $\upsilon_g = 1.8 \times 10^{-5}$ m$^2$/s. The two-phase surface tension is $\sigma = 0.072$ N/m. Grid independence is detailly examined under both quiescent and shear airflow conditions. Here, the Cahn number is introduced to quantify different gird resolutions, i.e., $Cn = \delta/D_0$, therein the interface thickness $\delta$ is fixed at 5 lattice units (lu), while $D_0$ varies among 50, 100, and 250 lu corresponding to three distinct resolutions: coarse ($Cn=0.1$), baseline ($Cn=0.05$), and finest ($Cn=0.02$). The multiphase system is characterized by two dimensionless numbers: the droplet Weber number and the airflow Reynolds number, defined as

$$We = \frac{\rho_l D_0 u_{d,z}^2}{\sigma}, \quad Re = \frac{D_0 u_{a,x}}{\upsilon_g} \tag{15}$$

First, for the benchmark case of normal droplet impact under quiescent conditions, as visualized in Fig. 2(a), the simulated droplet morphologies at the baseline resolution ($Cn=0.05$) show good agreement with the reported experimental images [58]. The dimensionless time is defined as the ratio of real time to the inertia-capillary time, i.e., $t^*=t/\sqrt{\rho_l D_0^3/(8\sigma)}$. The quantitative comparison is demonstrated in Fig. 2(b), where the time-evolution curves of the dimensionless droplet width ($W_x^*=W_x/D_0$) and spreading diameter ($D_x^*=D_x/D_0$) at the baseline resolution ($Cn=0.05$) are virtually indistinguishable from those obtained at the finest resolution ($Cn=0.02$), whereas the coarse resolution ($Cn=0.1$) yields relatively obvious deviations. To be more convincing, three cases with distinct impact velocities $u_{d,z}$ =0.55, 0.93, and 1.32 m/s ($We$ =10, 30, 60) are tested at the baseline resolution. Fig. 2(c) compares the temporal evolution of $W_x^*$ against our experimental data. The experimental setup, detailed in our previous works [59, 60], primarily consists of a visualized wind tunnel, a droplet generation module, and a high-speed data acquisition system. Obviously, the simulation captures the dynamic spreading and retraction features with high accuracy; moreover, as summarized in the inset table of Fig. 2(c), the obtained key metrics (including the maximum spreading width $W_{x,\max}^*$, total contact time $t_c^*$, and restitution coefficient $\epsilon_z$) exhibit consistency with the experimental values reported in previous literature for the similar scenario [23, 25]. These quantitative agreements confirm that the adopted LB model with the baseline resolution is capable of capturing the dominant behaviors of droplet impact, without loss of physical fidelity.

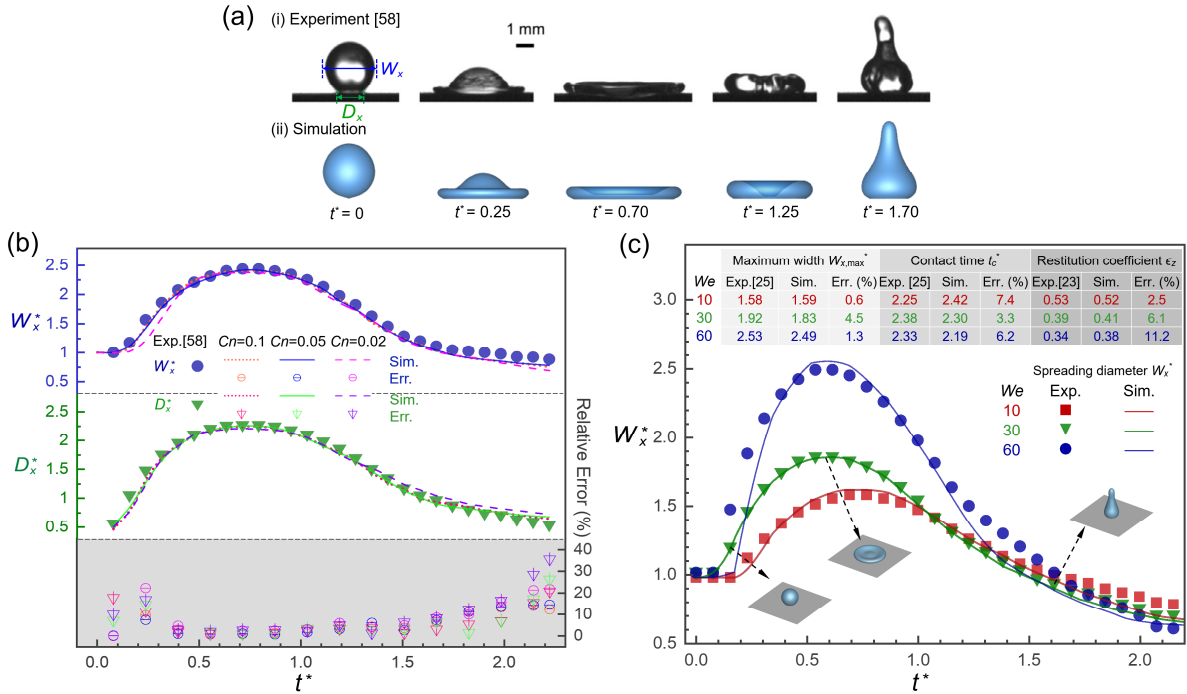

**Fig. 2.** Validation of normal droplet impact under quiescent conditions: (a) Comparison of the droplet morphological snapshots between the previous experiment [58] and simulation using the baseline resolution at $We$=44; (b) Grid independence verification of the dimensionless droplet width $W_x^*$ and streamwise spreading diameter $D_x^*$ at $We$=44 for three grid resolutions ($Cn$=0.1, 0.05, and 0.02); (c) Time-evolution of $W_x^*$ at $We$=10, 30, and 60. The inset table lists the quantitative comparison of maximum width $W_{x,\max}^*$, contact time $t_c^*$, and restitution coefficient $\epsilon_z$ against reported experimental data [23, 25].

Following the quiescent benchmark, the simulation accuracy is further validated against experiments under shear airflow conditions. For this validation case, the droplet impact velocity and bulk airflow velocity are set as $u_{d,z}$ =0.55 m/s ($We$ =10) and $u_{a,x}$ =4 m/s ($Re$ =560), respectively. Qualitatively, Fig. 3(a) illustrates the droplet morphology simulated with the baseline resolution. The numerical snapshots exhibit good agreement with the experimental images, correctly capturing the airflow-induced asymmetry. Meanwhile, Fig. 3(b) compares the time evolution of the dimensionless streamwise spreading diameter $D_x^*$ and slide displacement ($L_x^*=L_x/D_0$) across the three resolutions. The curves for the baseline and fine grids are virtually indistinguishable, demonstrating clear convergence, whereas the coarse grid also exhibits noticeable deviations. Notably, as listed in the inset table of Fig. 3(b), the relative errors of the obtained key metrics (maximum spreading diameter $D_{x,\max}^*$, total contact time $t_c^*$, and final displacement $L_{x,\max}^*$) against experimental measurements are relatively acceptable at the baseline resolution, which are close to those of the finest grid. This convergence indicates that the baseline resolution ($Cn$=0.05) is sufficient to reproduce the droplet dynamics under airflow, as further mesh refinement provides minimal enhancement in accuracy.

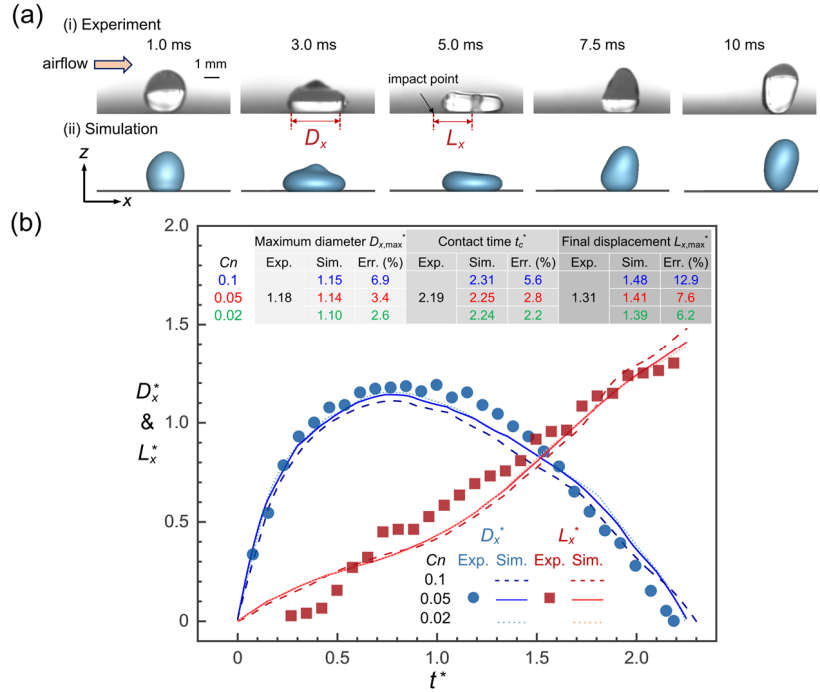

**Fig. 3.** Validation of droplet impact dynamics under shear airflow ($We$ =10 and $Re$ =560): (a) Comparison of the droplet morphological snapshots between the experiment and simulation using the baseline resolution; (b) Grid independence verification of the dimensionless streamwise spreading diameter $D_x^*$ and slide displacement $L_x^*$ for three grid resolutions ($Cn$=0.1, 0.05, and 0.02). The inset table lists the quantitative comparison of the maximum diameter $D_{x,\max}^*$, contact time $t_c^*$, and final displacement $L_{x,\max}^*$ against experimental data.

Thereby, at the chosen resolution of $Cn$=0.05, the factor $C_l$ =2.5×10$^{-5}$ m/lu is selected for all subsequent simulations as the conversion factor between macroscopic length (m) and lattice length unit (lu). Meanwhile, the mass and time conversion factors are set as $C_m$ =1.56×10$^{-11}$ kg/mu, $C_t$ =1.5×10$^{-6}$ s/ts, respectively, from which other derived units can be obtained. Unless otherwise specified, the simulation parameter setups in this paper are listed in Table 1, where the droplet properties could correspond to pure water or dilute saline at ambient temperature and pressure [61]. The computational domain is fixed at $6D_0\times3D_0\times3D_0$, with 54 million grid nodes.

Table 1. Simulation parameters with lattice units (LU) of length (lu), time (ts), and mass (mu).

| Variable | Macro value (IU) | LB value (LU) |
|---|---|---|
| droplet diameter $D_0$ | 2.5 mm | 100 lu |
| liquid-gas surface tension $\sigma$ | 0.072 N/m | 0.0104 mu/ts$^2$ |
| liquid density $\rho_l$ | 1000 kg/m$^3$ | 1 mu/lu$^3$ |
| gas density $\rho_g$ | 1.0 kg/m$^3$ | 0.001 mu/lu$^3$ |
| liquid viscosity $\upsilon_l$ | 1.0×10$^{-6}$ m$^2$/s | 0.0024 lu$^2$/ts |
| gas viscosity $\upsilon_g$ | 1.8×10$^{-5}$ m$^2$/s | 0.0432 lu$^2$/ts |
| static, receding, and advancing contact angles | $\theta_S$ =150°, $\theta_R$ =145°, $\theta_A$ =155° | |

## 3. Results and discussion

In this section, a systematic parameter study is conducted to quantitatively analyze the droplet impact dynamics on SHPS under shear airflow. The bulk airflow velocities vary within $u_{a,x}$ = 0, 1, 2, 3, 4, 5 m/s ($Re \approx$ 0, 140, 280, 420, 560, 700), and the initial droplet impact velocities are set to $u_{d,z}$ = 0.55, 0.76, 0.93, 1.08, 1.2, 1.32 m/s ($We \approx$ 10, 20, 30, 40, 50, 60). Notably, the investigated $We$ range ensures that the droplet remains in the complete rebound regime, avoiding the breakup or splashing over the critical threshold of high inertia impacts [62, 63]. Particular attention is directed toward the influence of airflow on contact-line characteristics (spreading factor, contact time, slide displacement, and final contact footprint) and detachment performance (velocity restitution coefficient and take-off angle). All simulations are accelerated using GPU-based parallel computing, with detailed efficiency metrics provided in the *Appendix*.

*3.1. Dynamic process of droplet impact*

As shown in Fig. 4, the presence of airflow significantly alters the morphological evolution of a droplet impinging on the SHPS. During the spreading stage from droplet to lamella ($t^* \leq$ 0.86), the curvature change of the liquid-gas interface perturbs the incoming airflow, and the airflow adheres to the liquid surface, leading to local reattachment over the spreading film. Driven by the inertial force of the reattached airflow, the rim of the pancake-shaped film exhibits an off-center forward motion, resulting in asymmetric deformation of the contact area along the streamwise direction. In the retraction stage ($t^* \geq$ 1.29), the low adhesion of the SHPS enables rapid recovery of the droplet's kinetic energy. In this case, the velocity gradient of the airflow in the $z$-direction imposes uneven aerodynamic shear on the rebounding droplet, thus it detaches from the surface with a thumb-like deformation and a deflected take-off angle. Meanwhile, a distinct slide displacement is observed on SHPS throughout the spreading and retraction process, due to kinetic energy absorbed from the expanding lamella and the windward side of the rebounding droplet.

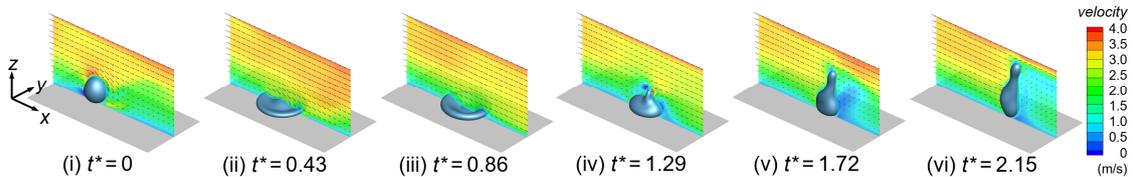

**Fig. 4.** Morphological evolution of a droplet impinging on SHPS ($We$=40, $Re$=560), along with the shear airflow velocity profiles.

The joint effects of airflow *Re* and impact *We* on the contact-line characteristics are discussed below. Figs. 5 (a), (b) present the droplet morphologies and airflow velocity profiles at the moments of maximum spreading and detachment for different *We* and *Re*, respectively. Correspondingly, Fig. 6 quantifies several key parameters of the contact surface on the SHPS, including the streamwise spreading factor $D_x^*$, transverse spreading factor $D_y^*$, deformation factor $D_d^*(=D_x^*-D_y^*)$, dimensionless contact area $A_c^*(=A_c/4\pi D_0^2)$, and slide displacement $L_x^*$; also, the transient evolutions of kinetic energy $E_k^*$, surface energy $E_s^*$, and viscous dissipation $E_v^*$, are demonstrated, whose values are normalized by the initial total energy and their calculations are detailed in Ref. [64].

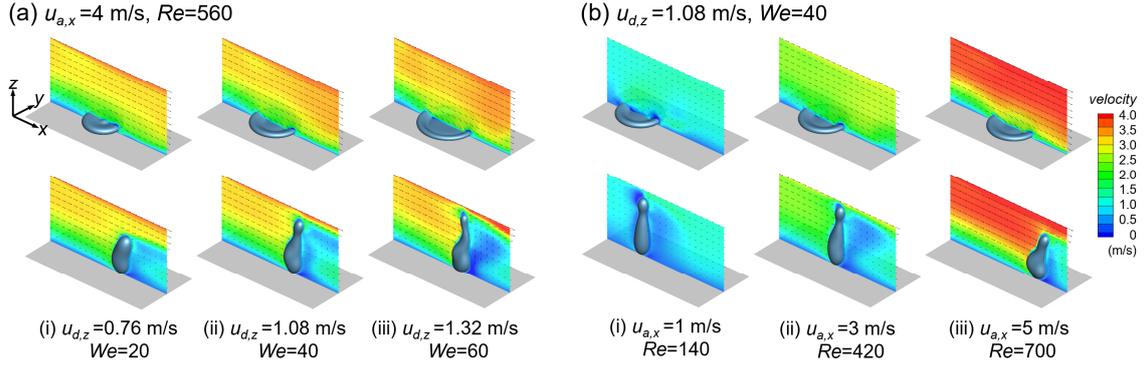

**Fig. 5.** Snapshots of droplet morphology and airflow velocity profiles at maximum spreading (upper row) and detachment moment (bottom row): (a) *Re*=560 with *We*=20, 40, and 60; (b) *We*=40 with *Re*=140, 420, and 700.

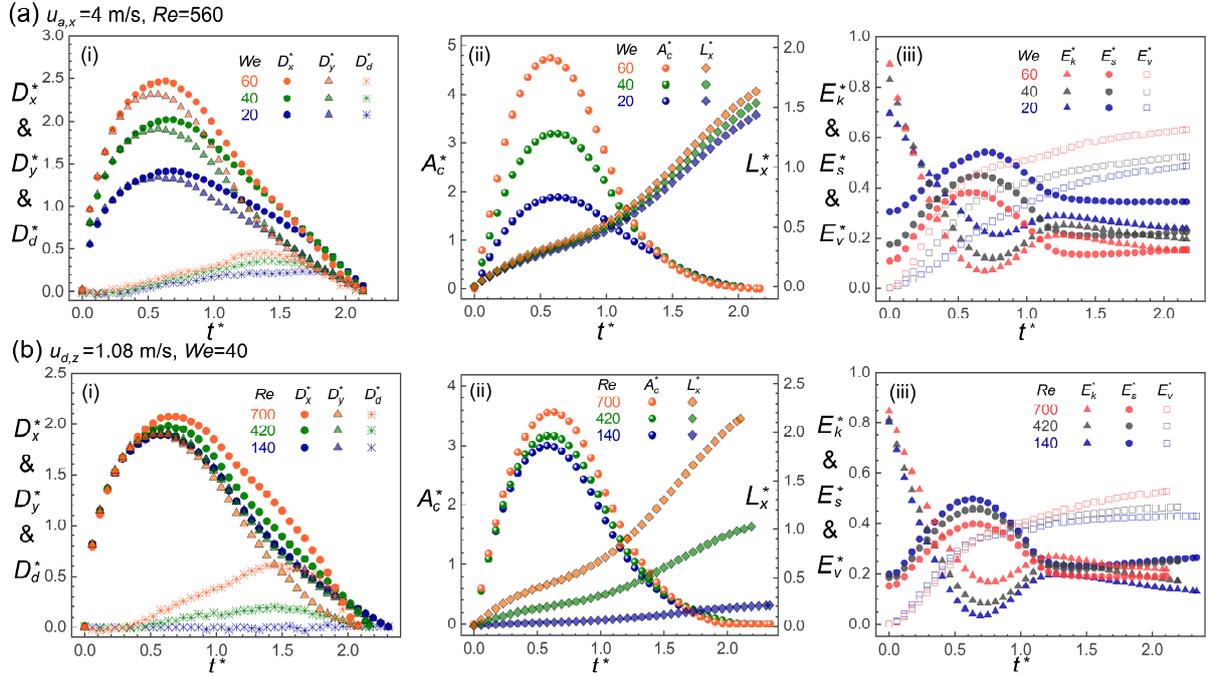

**Fig. 6.** Transient evolutions of (i) streamwise spreading factor $D_x^*$, transverse spreading factor $D_y^*$, and deformation factor $D_d^*$, (ii) dimensionless contact area $A_c^*$ and slide displacement $L_x^*$, and (iii) droplet kinetic energy $E_k^*$, surface energy $E_s^*$, and viscous dissipation $E_v^*$: (a) *Re*=560 with *We*=20, 40, and 60; (b) *We*=40 with *Re*=140, 420, and 700.

In the early stage ($t^*<0.5$), the curves of $D_x^*$ and $D_y^*$ almost overlap, with $D_d^*$ remaining close to zero, indicating that airflow exerts a negligible influence on the initial spreading. This is because the droplet retains a high residual kinetic energy while the expanding area is still small in this period, making the fraction of kinetic energy replenished by the airflow insignificant compared with the total kinetic energy [21].

During the fully spreading stage ($0.5<t^*<1$), where the droplet's initial kinetic energy is progressively converted into surface energy, the expanding lamella enlarges the liquid-gas interfacial area and strengthens their interaction. Therefore, the kinetic energy supply in the *x*-direction breaks spreading symmetry, causing divergence between $D_x^*$ and $D_y^*$ along with an increase in $D_d^*$, which indicates the contact area assumes an oval deformation in the streamwise direction. Meanwhile, airflow induces a horizontal motion of the contact line, leading to an increase of $L_x^*$. Specifically, as observed from the upper snapshots of Fig. 5(a), although a higher *We* makes the main body of the lamella within the viscous sublayer and weakens the kinetic energy absorption on the windward surface, the enlarged spreading area enhances the frictional drag of the reattached airflow over the droplet. Consequently, the droplet-airflow interaction is intensified to some extent by the impact inertia. As depicted in Fig. 6(a), this dynamic leads to a slight increase in both $D_d^*$ and $L_x^*$ as *We* increases. Moreover, the influence of *We* on the initial kinetic energy in the *z*-direction outweighs the contribution of spreading area variation to the kinetic energy supply in the *x*-direction during this stage [Fig. 6(a-iii)]; hence, *We* mainly determines the magnitudes of the two spreading factors and contact area. However, from the upper snapshots of Fig. 5(b), it is obvious that the increasing *Re* enhances the interaction between the expanding lamella and the reattached airflow within the viscous sublayer. In these cases, Fig. 6(b) demonstrates that $D_x^*$ and $L_x^*$ are greatly prolonged as *Re* increases, while $D_y^*$ remains independent of *Re* due to the preservation of transverse symmetry. Consequently, $D_d^*$, $L_x^*$, and $A_c^*$ are all positively correlated with the airflow *Re*.

Furthermore, in the later retraction stage ($t^*>1$), as the droplet accelerates toward detachment from the surface (with a increase in the slope of $L_x^*$), $A_c^*$ decreases sharply and $D_d^*$ reaches its peak. This occurs because the enlarged windward area of the rebounding droplet exposes it to higher-velocity airflow, as displayed in Fig. 4. Although both *We* and *Re* influence the contact behavior in this stage, the relative significance of their effects differs. Specifically, increasing *We* raises the bounce height, which enlarges the droplet's windward area [Fig. 5(a), bottom row]; however, the droplet root that occupies a larger proportion of its interfacial area, remains within the viscous sublayer, and viscous dissipation grows with *We* much more significantly than the energy supply [Fig. 6(a-iii)] [65]. Consequently, $D_d^*$ and $L_x^*$ show weak dependence on *We* in this stage. In contrast, increasing *Re* markedly enhances the aerodynamic shear acting on the rebounding droplet [Fig. 5(b), bottom row], resulting in a positive correlation of $D_d^*$ and $L_x^*$ with *Re*, along with stronger downstream deflection of the droplet apex. In these cases, although the droplet's retracting velocity increases, it is worth noting that the streamwise and transverse spreading diameters complete their retraction simultaneously under shear airflow (i.e., $D_x^*=D_y^*=0$ at detachment), which differs from the case of droplet impact on the moving surface [22, 66].

## 3.2. Droplet contact-line characteristics

### 3.2.1. Spreading factor

Fig. 7(a) illustrates the variations in the maximum streamwise and transverse spreading factors, $D^*_{x,\max}$ and $D^*_{y,\max}$ (abbreviated as $\beta_x$ and $\beta_y$) with respect to $We$ and $Re$. It is observed that $\beta_y$ remains largely independent of $Re$ across all $We$ cases. In contrast, $\beta_x$ is increased with $Re$, and this increment positively correlates with $We$, particularly at high $Re$. To account for the asymmetric spreading driven by airflow, a modified Weber number ($We^*$) is defined. This parameter represents the total effective kinetic energy contributed by both the vertical impact and the horizontal airflow replenishment, i.e.,

$$We^* = We\left[1+\left(\frac{u_{eff,x}}{u_{d,z}}\right)^2\right] \tag{16}$$

where $u_{eff,x} \propto u_{a,x}$ denotes the effective droplet velocity in the $x$-direction induced by the shear airflow. Considering that the aerodynamic acceleration acting on the droplet within the airflow layer decays exponentially, $u_{eff,x}$ can be approximated as [30, 67]

$$u_{eff,x} \doteq \frac{Re\left(1-e^{-t_{air}/\Gamma_{air}}\right)v_g}{D_0} \tag{17}$$

where $t_{air} = \sqrt{2h/g} = \sqrt{\sigma We/\rho_l D_0 g^2}$ is the droplet-airflow interaction time for a prescribed vertical $We$, and $\Gamma_{air} = 4\rho_l D_0^2/(3ReC_d\rho_g v_g)$ is the Stokes characteristic time, therein $C_d = 24(1+0.1Re^{0.75})/Re$ denotes the drag coefficient [68].

For droplet impact on SHPS within the inertial-capillary regime, Clanet et al.[12] established that the characteristic thickness $h$ of the pancake at maximum spreading is governed by the balance between inertial acceleration ($a_i \sim U_0^2/D_0$) and capillary deceleration ($a_c \sim \sigma/\rho_l h^2$). By substituting the standard kinetic energy with the effective kinetic energy, i.e., $U_0^2 = We^*\sigma/(\rho_l D_0)$, this balance yields $h \sim D_0(We^*)^{-1/2}$. Invoking volume conservation for the oval-shaped pancake, $D_0^3 \sim W_{x,\max}W_{y,\max}h$, the scaling for the maximum spreading widths is derived as

$$W_{x,\max}W_{y,\max} \sim D_0^2(We^*)^{1/2} \tag{18}$$

Notably, as illustrated in Fig. 1(b), the droplet's maximum spreading width $W$ and the actual contact diameter $D$ on the solid surface are not identical due to the bulging rim. For the regime investigated herein, their values can be converted using the relation proposed by Hu et al. [58], i.e., $W/D \propto (We^*)^{-3/5} + A$, where $A$ is a fitting constant. Coupling this relation with Eq. (18), the composite scaling law for the dimensionless contact area $\beta_x\beta_y$ is obtained as

$$\beta_x\beta_y \propto \frac{(We^*)^{1/2}}{(We^{*-3/5}+A)^2} \tag{19}$$

To determine the constant $A$ and validate this scaling framework, the fit of Eq. (19) is performed against all simulation data across varying $Re$ and $We$ cases (corresponding to $We^*$=10~72). As depicted in Fig. 7(b), the simulation data consistently collapses onto the theoretical scaling curve with a high determination coefficient of $R^2$=0.983, and $A$ is optimally fitted as 0.21. As shown in the inset pairplot, the relative deviation between theoretical and simulated results is less than 10%, validating the reliability of the composite scaling relation.

Then, the explicit airflow influence on the streamwise spreading is decoupled. Since the spreading dynamics in the $y$-direction are orthogonal to the airflow, $\beta_y$ is almost independent of $Re$ [as demonstrated in Fig. 7(a)]. Consequently, $\beta_y$ across all $Re$ cases basically follows the same scaling at $Re=0$ where $We^*=We$, expressed as

$$\beta_y \doteq \sqrt{(\beta_x \beta_y)|_{Re=0}} \propto \frac{We^{1/4}}{We^{-3/5}+A} \quad (20)$$

Thus, the streamwise spreading factor $\beta_x$ prolonged by airflow can be quantified as

$$\beta_x = \frac{\beta_x \beta_y}{\beta_y} \propto \frac{We^{*1/2}(We^{-3/5}+A)}{We^{1/4}(We^{*-3/5}+A)^2} \quad (21)$$

Figs. 7(c) and (d) clearly demonstrate that $\beta_x$ and $\beta_y$ align perfectly with the derived scaling relations of Eqs. (20) and (21), respectively ($R^2>0.98$, relative deviations<10%). Moreover, it is evident that the product of each pre-factors excellently matches the pre-factor of $\beta_x \beta_y$ (i.e., 0.2369×0.2372≈0.0562). This provides a solid theoretical foundation for subsequent prediction of the contact surface morphodynamics.

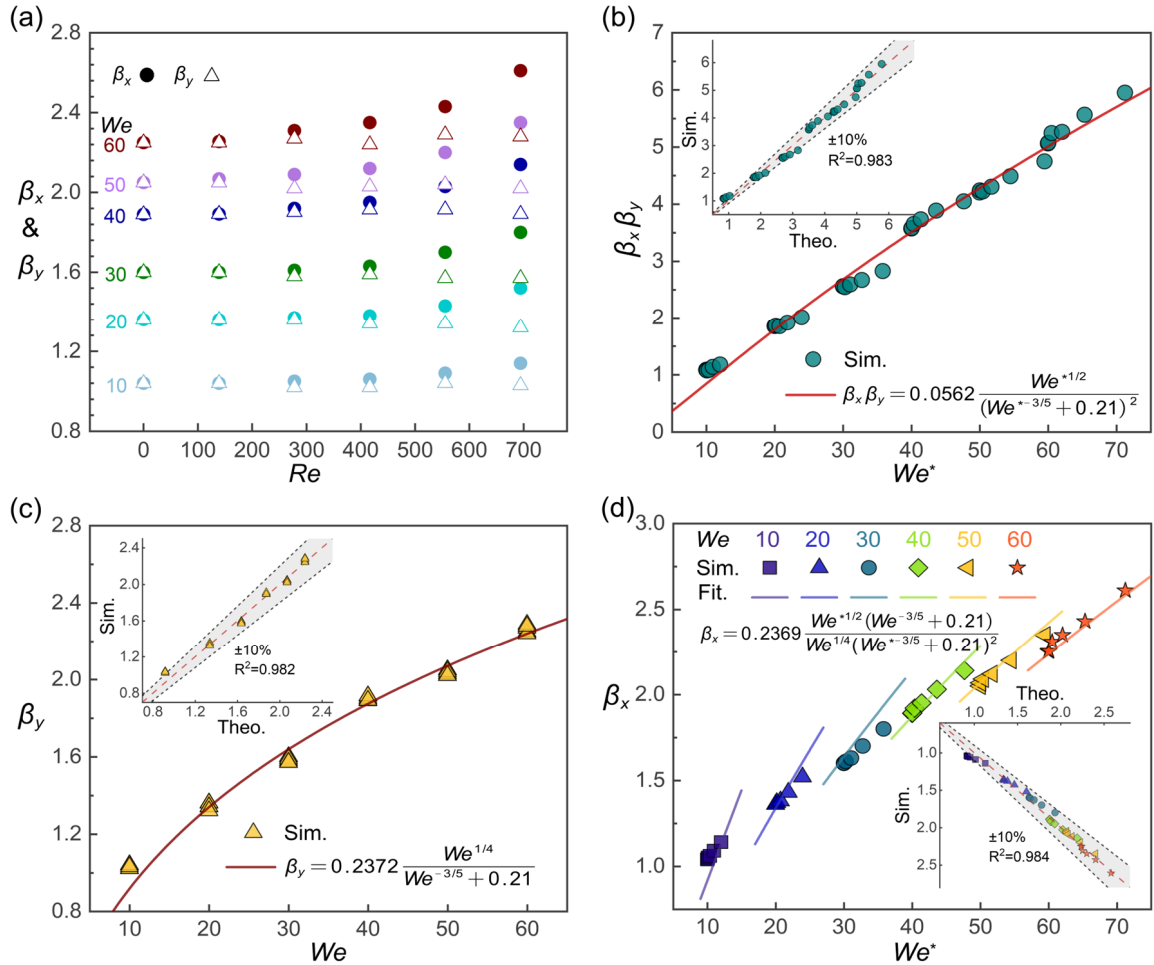

**Fig. 7.** (a) Maximum streamwise and transverse spreading factors ($\beta_x$, solid circles, and $\beta_y$, hollow triangles) at various $Re$ and $We$; (b) Comparison between the simulation data and theoretical relation of Eq. (19) for the dimensionless contact area ($\beta_x \beta_y$), along with the inset pairplot for their deviations; (c) and (d) Validations of the decoupled scaling laws of Eqs. (20) and (21) for $\beta_y$ and $\beta_x$, respectively.

*3.2.2. Slide displacement and contact footprint*

Since the droplet motion from initial contact to final detachment under shear airflow is accompanied by continuous horizontal sliding, the final contact footprint is no longer adequately represented by the maximum spreading area $A_{c,\max}$ in Fig. 6(ii), which is only valid for the no-sliding case in quiescent air [69]. Instead, it should be defined as the Boolean union of the transient contact area $A_c$ over the entire contact duration, denoted as $A_U$. This metric integrates the combined effects of the morphodynamic spreading factor, contact time, and slide displacement.

Based on our simulations and previous experimental observations [70, 71], the geometric composition of $A_U$ depends on the magnitude of the slide displacement $L_s$. As schematically illustrated in Fig. 8(a, left), if the droplet fails to slide beyond its maximum spreading boundary during the retraction phase (i.e., a minor slide displacement satisfying $L_s t_r^*/t_c^* \leq D_{x,\max}/2$), $A_U$ remains equivalent to the elliptical area at maximum spreading. Considering that across all simulated cases presented in Fig. 8(b), the retraction time $t_r^*$ accounts for approximately 70% of the total contact time $t_c^*$, the slide displacement during retraction can thus be estimated as $L_s t_r^*/t_c^* \approx 0.7 L_s$. However, under strong airflow [Fig. 8(a), right], the droplet slides out of the maximum spreading region. In this scenario, $A_U$ is geometrically characterized by two adjoined semi-ellipses with distinct eccentricities: one half preserves the original profile of the maximum spreading, while the other is further elongated by the sliding sweep.

Consequently, evaluating $A_U$ inherently requires determining the slide displacement $L_s$ [72]. For droplet motion on SHPS, the contact-line pinning friction is negligible compared to the aerodynamic forces [73-75]. The sliding motion is predominantly propelled by the dynamic pressure of the airflow, $P_{dyn} = \rho_g u_{in,x}^2/2 \propto Re^2$, and the total aerodynamic thrust is fundamentally composed of two components [76]: the form drag ($F_{form}$) acting on the frontal windward area, and the viscous skin friction ($F_{skin}$) acting on the exposed upper film. Therein, the form drag is expressed as $F_{form} \sim C_d P_{dyn} A_{front}$, where $C_d$ is the form drag coefficient and $A_{front} \sim W_y h$. Recalling the characteristic thickness $h \sim D_0^3/(W_x W_y)$ derived from the volume conservation of the pancake, the relation $F_{form} \propto Re^2/\beta_x$ is mathematically yielded. Moreover, driven by the shear stress on the exposed upper planar area $A_{top} \sim W_x W_y \propto \beta_x \beta_y$ with a friction coefficient $C_f$, the viscous skin friction is obtained as $F_{skin} \sim C_f P_{dyn} A_{top} \propto Re^2 \beta_x \beta_y$. According to Newton's second law, the slide displacement is $L_s \sim a_x t_c^2/2$, where the sliding acceleration is $a_x \sim (F_{form}+F_{skin})/(\rho_l D_0^3)$, and the total contact time $t_c$ can be considered as a constant [Fig. 8(b)]. Eventually, grouping all constant fluid properties and coefficients into two empirical pre-factors, $C_{form}$ and $C_{skin}$, the relation for the dimensionless slide displacement $L_s^* = L_s/D_0$ can be written as

$$L_s^* = Re^2 \left( \frac{C_{form}}{\beta_x} + C_{skin} \beta_x \beta_y \right) \tag{22}$$

Fig. 8(c) illustrates the variation of $L_s^*$ across different *We* and *Re*. The derived scaling law achieves an excellent determination coefficient of R$^2$=0.944, with relative deviations from the simulation data within 15%. It is observed that $L_s^*$ exhibits a strong positive correlation with $Re$ but only a weak dependence on $We$. This trend can be qualitatively elucidated by Eq. (22): as the droplet severely flattens at higher $We$, the loss in form drag (due to the reduced frontal profile $\propto \beta_x^{-1}$) is additionally compensated by the increase of skin friction (due to the expanded area $\propto \beta_x \beta_y$), thus leading to $L_s^*$ within an airflow-dominated state.

Thereby, with $L_s$ theoretically closed, the final contact footprint $A_U$ can be systematically evaluated, whose dimensionless value is given by the following piecewise function

$$A_U^* = \frac{A_U}{\frac{\pi}{4} D_0^2} = \begin{cases} \beta_x \beta_y, & \text{if } \beta_x \geq 1.4 L_s^*, \\ 0.5 \beta_x \beta_y + 0.7 L_s^* \beta_y, & \text{otherwise.} \end{cases} \quad (23)$$

As shown in Fig. 8(d), the predicted results from Eq. (23) exhibit excellent agreement with the simulations, demonstrating a high accuracy with $R^2=0.969$ and quantitative deviations within 15%. Notably, the curves remain relatively flat at low $Re$. In this regime, the condition $\beta_x \geq 1.4 L_s^*$ is satisfied, indicating that the contact footprint is predominantly governed by the impact deformation; however, once the airflow velocity exceeds a critical threshold (high $Re$), the droplet undergoes obvious sliding, so that the curves exhibit a inflection and increases rapidly by up to 80% compared with $Re=0$.

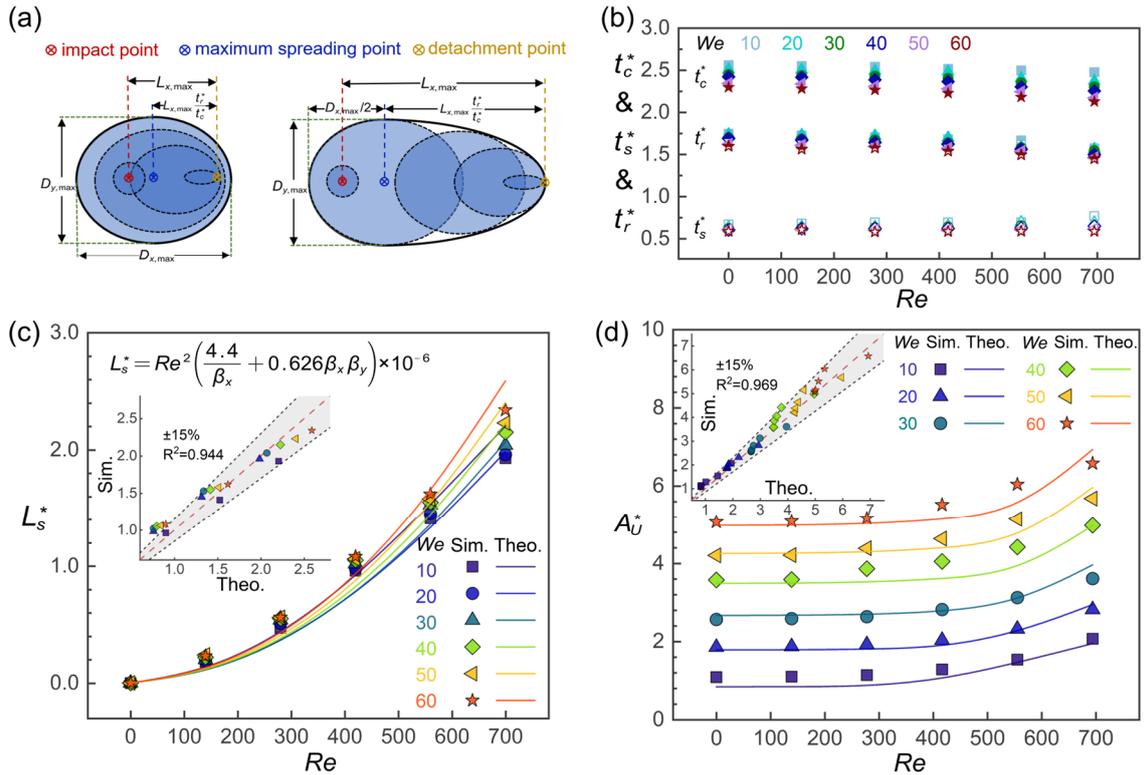

**Fig. 8.** (a) Schematic definition of the final contact footprint $A_U$ of a droplet under shear airflow, distinguishing between minor and strong sliding regimes; (b) Dimensionless contact time at various $Re$ and $We$, including the total time $t_c^*$, the duration of spreading $t_s^*$, and retraction $t_r^*$, where the sample colors correspond to differernt $We$ cases; (c) and (d) Comparisons between the theoretical prediction and simulation results for the dimensionless slide displacement $L_s^*$ and contact footprint $A_U^*$, along with the inset pairplots for their deviations.

In summary, both the numerical and theoretical results confirm that the droplet contact-line characteristics on the SHPS are governed by the coupled effects of *We* and *Re*. While the presence of airflow breaks the symmetry of the spreading dynamics, its effect on the contact time is relatively weak. Notably, due to the extremely low adhesion of the SHPS, the slide displacement exhibits a high sensitivity to aerodynamic shear. This kinematic sliding induces a substantial expansion of the final contact footprint, which can significantly influence the actual heat and mass transfer processes between the droplet and the solid surface [59, 60].

*3.3. Droplet detachment characteristics*

*3.3.1. Energy partition*

As shown in Fig. 6(iii), the presence of airflow also alters the energy evolution during droplet impact on the SHPS. The droplet kinetic energy is replenished through momentum transfer from the airflow, which largely drives the horizontal sliding. Meanwhile, a part of the energy is converted into surface energy during the spreading stage, thereby enlarging the lamella contact area. Because of this extra kinetic energy supply from the airflow, the proportion of surface energy decreases with increasing airflow intensity. At the same time, the non-uniform aerodynamic shear causes irregular thumb-like deformations in the late retraction stage, resulting in increased viscous dissipation.

It is noteworthy that the droplet energy at the detachment instant receives particular attention, as it determines the subsequent dynamics of the rebounding droplet, e.g., secondary rebound and transport height [77]. Fig. 9(a) shows the proportions of kinetic energy ($E_k'$), surface energy ($E_s'$), and viscous dissipation ($E_v'$) at this moment for various *We* and *Re* conditions. As both *We* and *Re* increase simultaneously, $E_v'$ increases whereas $E_s'$ decreases. In addition, although $E_k'$ increases with *Re*, the growth rate gradually weakens with higher *We*. This is because the initial kinetic energy provided by the vertical impact velocity is much greater than the energy replenished by the airflow.

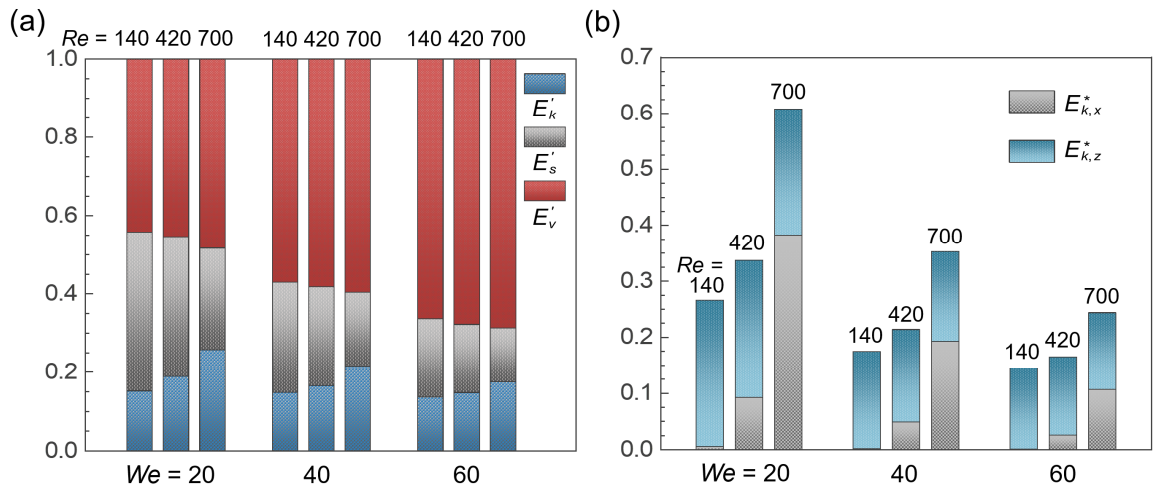

**Fig. 9.** Droplet energy partition of the detachment moment at various *Re* and *We*: (a) Proportions of kinetic energy $E_k'$, surface energy $E_s'$, and viscous dissipation $E_v'$; (b) Ratios of vertical and streamwise kinetic energies to the initial total kinetic energy, denoted as $E_{k,z}^*$ and $E_{k,x}^*$, respectively.

Neglecting the transverse kinetic energy in the $y$-direction, Fig. 9(b) demonstrates kinetic energies at detachment of the vertical direction $E_{k,z}^*$ and streamwise direction $E_{k,x}^*$, normalized by the initial kinetic energy. Obviously, the two energies exhibit different sensitivities to $We$ and $Re$: $E_{k,z}^*$ is weakly dependent of $Re$ but decreases markedly with $We$, whereas $E_{k,x}^*$ increases with $Re$ and decreases with $We$. Consequently, the total droplet kinetic energy at the onset of take-off [indicated by the total column height in Fig. 9(b)] is jointly controlled by $We$ and $Re$. To quantitatively characterize the droplet detachment behavior induced by airflow, particularly the recovery of kinetic energy, two directional velocity restitution coefficients are defined as [78]

$$\epsilon_z = u_{b,z}/u_{d,z} \doteq \sqrt{E_{k,z}^*}, \quad \epsilon_x = u_{b,x}/u_{d,z} \doteq \sqrt{E_{k,x}^*} \tag{24}$$

where $u_{b,x}$ and $u_{b,z}$ denote the streamwise and vertical bouncing velocities, respectively.

*3.3.2. Restitution coefficient and take-off angle*

Fig. 10(a) presents the vertical restitution coefficient $\epsilon_z$ for all $Re$ and $We$ cases. As indicated by the energy partition analysis in Sec. 3.3.1, the presence of airflow enhances kinetic energy replenishment, necessitating a coupled dependence of $\epsilon_z$ on both $We$ and $Re$. For droplet impact in quiescent air, Aria et al. [77] established the raw relation of $\epsilon_z = aWe^{-1/4}$, which originates from the 1/2-scaling law for maximum stored surface energy ($E_{s,\max} \propto We^{1/2}$) [12, 23]. Since the vertical rebound kinetic energy upon detachment from the SHPS is proportional to $E_{s,\max}$, the vertical bouncing velocity follows $u_{b,z}^2 \propto E_{s,\max}$ [23, 79]. By substituting the kinetic energy with the modified relation, i.e., $E_{s,\max} \propto We^{*1/2}$ according to Eq. (18), a generalized scaling law for $\epsilon_z$ is derived as

$$\epsilon_z = aWe^{*1/4}We^{-1/2} \tag{25}$$

As illustrated in Fig. 10(a), fitting this analytical model against our simulation data yields an optimal pre-factor of $a = 1.025$, with relative deviation less than 10% and $R^2 = 0.933$. Notably, this fitted value closely aligns with previously reported constants $a = 0.94$ or $1.1$ [23, 77]. Besides, in the absence of airflow ($We^* = We$), this relation reduces to the raw scaling $\epsilon_z = aWe^{-1/4}$, consistent with the finding of Aria et al. [77].

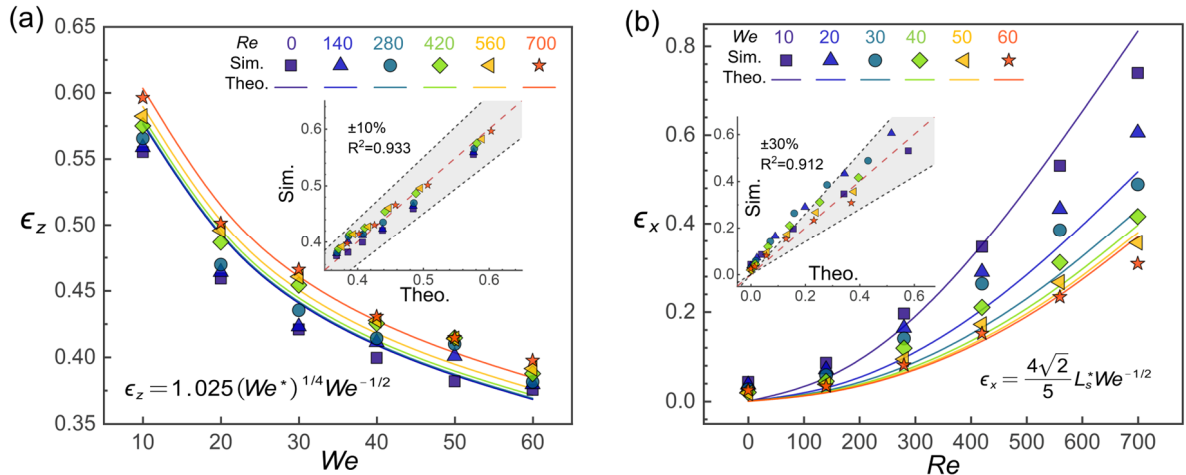

**Fig. 10.** Comparisons between the theoretical prediction and simulation results for the directional velocity restitution coefficients, along with the inset pairplots for their deviations: (a) Vertical velocity restitution coefficient $\epsilon_z$; (b) Streamwise velocity restitution coefficient $\epsilon_x$.

Furthermore, owing to the low friction of the SHPS, the energy loss during the horizontal sliding is negligible [69, 80]. Thus, the streamwise rebound velocity can be approximated by the average sliding velocity, i.e., $u_{b,x} \sim L_s/t_c$, where $L_s$ is obtained from Eq. (22). As previously demonstrated in Fig. 8(b), the contact time exhibits only a weak dependence on $Re$ and $We$, therefore, it can be treated as the theoretical constant here, i.e., $t_c^* \doteq 2.5$ [14]. Integrating $L_s$ and $t_c$, the streamwise velocity restitution coefficient $\epsilon_x$ is formulated as

$$\epsilon_x = \frac{2.5 L_{x,\max}^* D_0 / \left(\pi\sqrt{\rho_l D_0^3/(8\sigma)}\right)}{\sqrt{(\sigma We)/(\rho_l D_0)}} = \frac{4\sqrt{2}}{5} L_s^* We^{-1/2} \qquad (26)$$

As shown in Fig. 10(b), the predicted values from Eq. (26) well match the simulated data, achieving $R^2$=0.912 with relative deviations within 30%. Clearly, $\epsilon_x$ exhibits a strong positive correlation with $Re$ but a negative correlation with $We$, which suggest that although strong aerodynamic shear at high $Re$ drives horizontal momentum, massive impact inertia at high $We$ inherently suppresses the relative tangential acceleration. This conclusion is consistent with the trend of $E_{k,x}^*$ in Fig. 9(b).

Subsequently, based on the theoretical expressions of the vertical and streamwise velocity restitution coefficients [Eqs. (25) and (26)], Two quantitative parameters for evaluating droplet detachment behaviors are calculated: the total velocity restitution coefficient $\epsilon = \sqrt{\epsilon_x^2 + \epsilon_y^2}$ and the take-off angle $\alpha=\arctan(\epsilon_z/\epsilon_x)$. Fig. 11 compares these theoretical predictions with the simulation results, where the relative deviations of $\epsilon$ and $\alpha$ are within 15% and 20%, respectively. It is evident that the proposed model accurately captures the coupled influence of airflow $Re$ and impact $We$ on the droplet dynamics of the rebound stage. Overall, these results prove that our scaling theory is highly reliable for predicting droplet detachment behaviors under shear airflow.

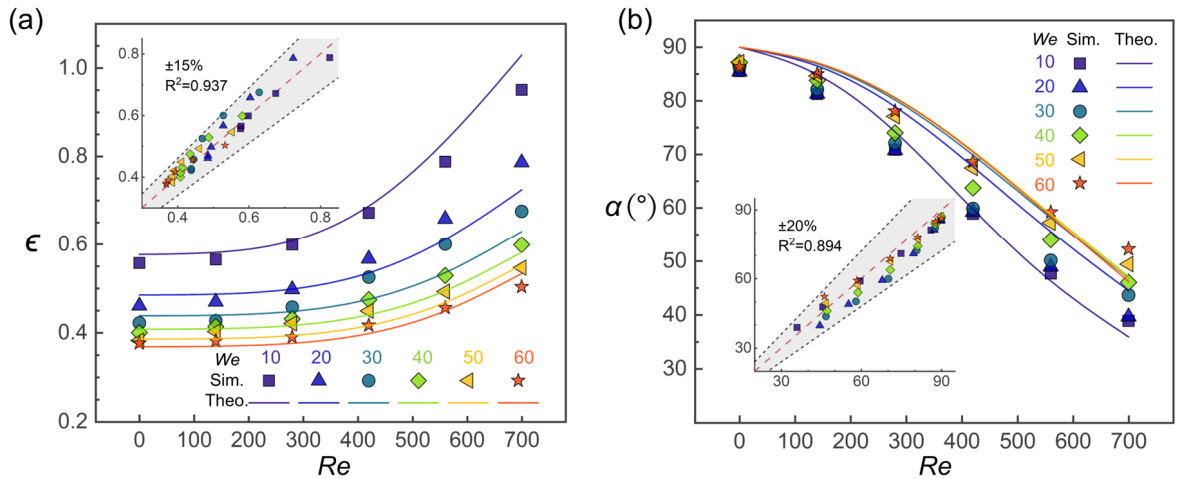

**Fig. 11.** Comparisons between the theoretical prediction and simulation results for droplet detachment characteristics, along with the inset pairplots for their deviations: (a) Total velocity restitution coefficient $\epsilon$; (b) Take-off angle $\alpha$.

## 4. Conclusions

In this work, a three-dimensional pseudopotential multiphase lattice Boltzmann model was employed to simulate high-density-ratio liquid-gas systems. The model incorporates a non-orthogonal multiple-relaxation-time scheme to enhance numerical stability and a contact angle hysteresis window to accurately reproduce dynamic wetting behavior on SHPS. Validated against experimental data, the model was applied to investigate droplet impact dynamics under shear airflow, leading to the following conclusions:

1. Shear airflow strongly couples with the morphological evolution of the droplet. The kinetic energy replenished by the airflow breaks the symmetry of the impact process, which is evidenced by the elliptical deformation of the contact area due to airflow reattachment on the expanding lamella, the unsteady contact-line sliding during both spreading and retraction phases, and a thumb-like detachment shape accompanied by a deflected take-off angle induced by aerodynamic shear.

2. Shear airflow increases the streamwise spreading and notably enlarges the final contact footprint by up to 80% due to horizontal sliding. Incorporating the airflow contribution on kinetic energy, a modified Weber number $We^* \sim f(We, Re)$ is introduced to reconstruct a composite scaling law to characterize the maximum spreading state. Furthermore, through the dimensional analysis of the aerodynamic forces, a theoretical relation for the slide displacement is obtained, which is subsequently employed to formulate a predictive relation for the final contact footprint. These theoretical models effectively resolve the contact-line characteristics nonlinearly-coupled between airflow effect and impact inertia.

3. Grounded in the analysis of energy partition at detachment, a refined scaling law for the vertical velocity restitution coefficient is derived to describe the recovered kinetic energy under different $We$ and $Re$ conditions. Additionally, a theoretical relation for the streamwise velocity restitution coefficient is formulated using the sliding velocity approximation. By integrating these decoupled directional components, the total velocity restitution coefficient and the trajectory take-off angle are quantitatively predicted, with relative deviations compared to simulation data less than 15% and 20%, respectively.

To sum up, the results of this study have the potential to offer theoretical insights into the evaluation of droplet dynamics behaviors under real-world airflow environments, as well as the optimization of functional surface designs.

**Acknowledgements**

This research is funded by the Natural Science Foundation of China (Grant No. 52176079 and No. 52376061). We appreciate Dr. Shaotong Fu (NVIDIA Beijing) for his technical support in GPU-accelerated computing.

**Appendix**

The discrete velocity set for the D3Q19 lattice stencil and the corresponding non-orthogonal transformation matrix are given by

$$\mathbf{e}_i = \begin{bmatrix} 0 & 1 & -1 & 0 & 0 & 0 & 0 & 1 & -1 & 1 & -1 & 1 & -1 & 1 & -1 & 0 & 0 & 0 & 0 \\ 0 & 0 & 0 & 1 & -1 & 0 & 0 & 1 & 1 & -1 & -1 & 0 & 0 & 0 & 0 & 1 & -1 & 1 & -1 \\ 0 & 0 & 0 & 0 & 0 & 1 & -1 & 0 & 0 & 0 & 0 & 1 & 1 & -1 & -1 & 1 & 1 & -1 & -1 \end{bmatrix} \quad (A1)$$

$$\mathbf{M} = \begin{bmatrix} 1 & 1 & 1 & 1 & 1 & 1 & 1 & 1 & 1 & 1 & 1 & 1 & 1 & 1 & 1 & 1 & 1 & 1 & 1 \\ 0 & 1 & -1 & 0 & 0 & 0 & 0 & 1 & -1 & 1 & -1 & 1 & -1 & 1 & -1 & 0 & 0 & 0 & 0 \\ 0 & 0 & 0 & 1 & -1 & 0 & 0 & 1 & 1 & -1 & -1 & 0 & 0 & 0 & 0 & 1 & -1 & 1 & -1 \\ 0 & 0 & 0 & 0 & 0 & 1 & -1 & 0 & 0 & 0 & 0 & 1 & 1 & -1 & -1 & 1 & 1 & -1 & -1 \\ 0 & 0 & 0 & 0 & 0 & 0 & 0 & 1 & -1 & -1 & 1 & 0 & 0 & 0 & 0 & 0 & 0 & 0 & 0 \\ 0 & 0 & 0 & 0 & 0 & 0 & 0 & 0 & 0 & 0 & 0 & 1 & -1 & -1 & 1 & 0 & 0 & 0 & 0 \\ 0 & 0 & 0 & 0 & 0 & 0 & 0 & 0 & 0 & 0 & 0 & 0 & 0 & 0 & 0 & 1 & -1 & -1 & 1 \\ 0 & 1 & 1 & 1 & 1 & 1 & 1 & 2 & 2 & 2 & 2 & 2 & 2 & 2 & 2 & 2 & 2 & 2 & 2 \\ 0 & 1 & 1 & -1 & -1 & 0 & 0 & 0 & 0 & 0 & 0 & 1 & 1 & 1 & 1 & -1 & -1 & -1 & -1 \\ 0 & 1 & 1 & 0 & 0 & -1 & -1 & 1 & 1 & 1 & 1 & 0 & 0 & 0 & 0 & -1 & -1 & -1 & -1 \\ 0 & 0 & 0 & 0 & 0 & 0 & 0 & 1 & -1 & 1 & -1 & 0 & 0 & 0 & 0 & 0 & 0 & 0 & 0 \\ 0 & 0 & 0 & 0 & 0 & 0 & 0 & 0 & 0 & 0 & 0 & 1 & -1 & 1 & -1 & 0 & 0 & 0 & 0 \\ 0 & 0 & 0 & 0 & 0 & 0 & 0 & 1 & 1 & -1 & -1 & 0 & 0 & 0 & 0 & 0 & 0 & 0 & 0 \\ 0 & 0 & 0 & 0 & 0 & 0 & 0 & 0 & 0 & 0 & 0 & 1 & 1 & -1 & -1 & 0 & 0 & 0 & 0 \\ 0 & 0 & 0 & 0 & 0 & 0 & 0 & 0 & 0 & 0 & 0 & 0 & 0 & 0 & 0 & 1 & -1 & 1 & -1 \\ 0 & 0 & 0 & 0 & 0 & 0 & 0 & 0 & 0 & 0 & 0 & 0 & 0 & 0 & 0 & 1 & 1 & -1 & -1 \\ 0 & 0 & 0 & 0 & 0 & 0 & 0 & 1 & 1 & 1 & 1 & 0 & 0 & 0 & 0 & 0 & 0 & 0 & 0 \\ 0 & 0 & 0 & 0 & 0 & 0 & 0 & 0 & 0 & 0 & 0 & 1 & 1 & 1 & 1 & 0 & 0 & 0 & 0 \\ 0 & 0 & 0 & 0 & 0 & 0 & 0 & 0 & 0 & 0 & 0 & 0 & 0 & 0 & 0 & 1 & 1 & 1 & 1 \end{bmatrix} \quad (A2)$$

In the pseudopotential multiphase LB model, the liquid-gas interfacial tension $\sigma$ can be adjusted via the numerical parameter $\kappa$ in Eq. (9). Fig. A1 shows the simulated values of $\sigma$, which are obtained from the pressure difference between the inside and outside of a static droplet according to Laplace's law, i.e., $\Delta p = 2\sigma/R_0$. For the water-air system considered in this study, $\kappa$ is fixed at 0.02, and the corresponding $\sigma$ is 0.0104 mu/ts$^2$.

Direct numerical simulations of the three-dimensional multiphase problems are computationally expensive. In the present study, the computational domain contained about 54 million lattice nodes, and thus the GPU parallelization is adopted to accelerate the computations, with the code compiled based on the open-source Python *Warp* platform. Initialization, data input, and output are executed on the CPU host, whereas the key steps (collision, streaming, and macroscopic variable calculation) are performed on the GPU device. As shown in Fig. A2, GPU computing significantly reduced the simulation time, with the NVIDIA 4090D achieving speedups of approximately one order of magnitude compared with server-grade CPUs and two orders of magnitude compared with consumer-grade CPUs.

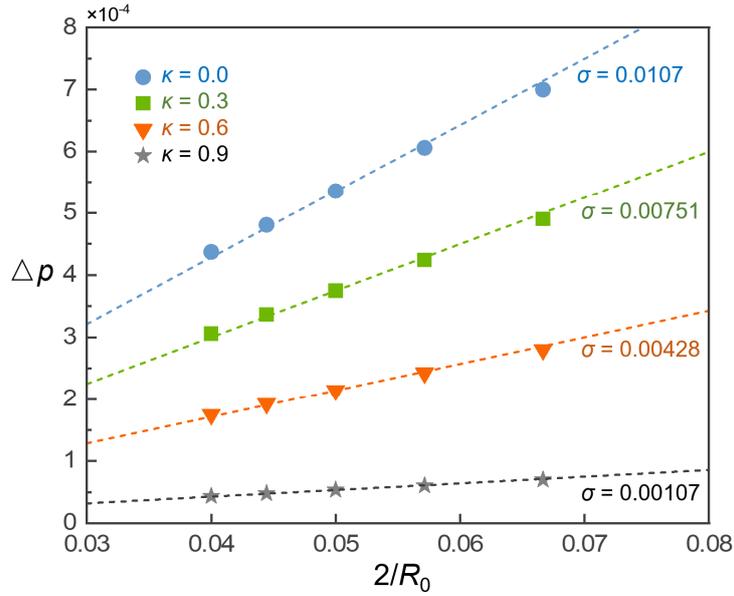

**Fig. A1.** Laplace test results for a static droplet.

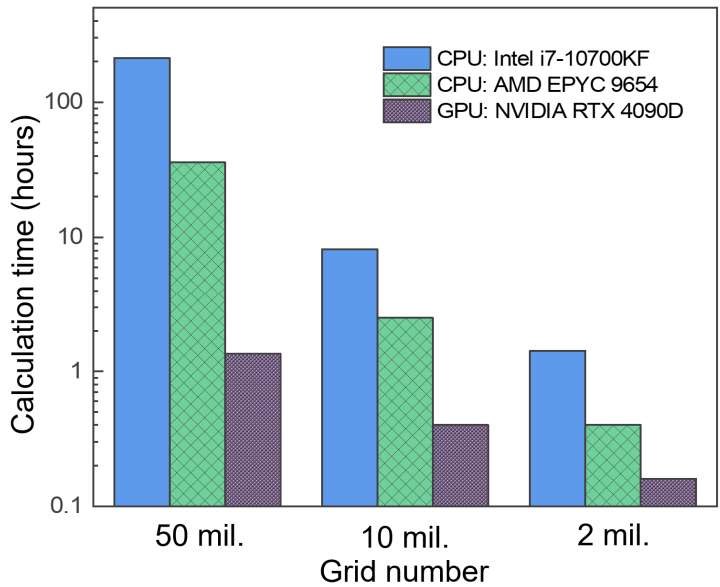

**Fig. A2.** Comparison of computation time between GPU and CPU parallel computing.